\documentclass[twocolumn,twocolappendix]{aastex631}

\newcommand\soutm{\bgroup\markoverwith{\textcolor{black}{\rule[0.5ex]{2pt}{0.8pt}}}\ULon}

\shorttitle{The fraction of quenched galaxies}
\shortauthors{Park et al.}
\graphicspath{{./}{figures/}}

\begin{document}

\title{Galaxy Quenching with Mass Growth History of Galaxy Groups and Clusters: The Importance of Post-Processing}

\correspondingauthor{Kyungwon Chun}
\email{kwchun@kasi.re.kr}

\author{So-Myoung Park}
\affiliation{Korea Astronomy and Space Science Institute, 776 Daedeok-daero, Yuseong-gu, Daejeon 34055, South Korea}

\author{Kyungwon Chun}
\affiliation{Korea Astronomy and Space Science Institute, 776 Daedeok-daero, Yuseong-gu, Daejeon 34055, South Korea}

\author{Jihye Shin}
\affiliation{Korea Astronomy and Space Science Institute, 776 Daedeok-daero, Yuseong-gu, Daejeon 34055, South Korea}

\author{Hyunjin Jeong}
\affiliation{Korea Astronomy and Space Science Institute, 776 Daedeok-daero, Yuseong-gu, Daejeon 34055, South Korea}

\author{Joon Hyeop Lee}
\affiliation{Korea Astronomy and Space Science Institute, 776 Daedeok-daero, Yuseong-gu, Daejeon 34055, South Korea}

\author{Mina Pak}
\affiliation{Korea Astronomy and Space Science Institute, 776 Daedeok-daero, Yuseong-gu, Daejeon 34055, South Korea}
\affiliation{School of Mathematical and Physical Sciences, Macquarie University, Sydney, NSW 2109, Australia}
\affiliation{ARC Centre of Excellence for All Sky Astrophysics in 3 Dimensions (ASTRO 3D), Australia}

\author{Rory Smith}
\affiliation{Universidad T\'{e}cnica Federico Santa Mar\'{i}a, 3939 Vicu\~{n}a Mackenna, San Joaqu\'{i}n, Santiago 8940897, Chile}

\author{Jae-Woo Kim}
\affiliation{Korea Astronomy and Space Science Institute, 776 Daedeok-daero, Yuseong-gu, Daejeon 34055, South Korea}



\begin{abstract}
We investigate the fraction of quenched satellite galaxies in host galaxy groups and clusters using TNG300 in the IllustrisTNG cosmological magnetohydrodynamical simulations.
Simulations show that most satellites are quenched after they fall into their final hosts: post-processing is a more dominant mechanism of galaxy quenching than pre-processing.
We find the fraction of quenched satellites at $z=0$ increases with host mass, which implies that more massive hosts have higher quenching efficiency because more massive hosts have more massive groups infalling.
Furthermore, we find that hosts that have many early-infall satellites show a higher fraction of quenched satellites at $z=0$ than those having many late-infall satellites, which results in a scatter of the quenched fraction of satellites in a given mass range of hosts at $z=0$.
Our results highlight the significance of the mass of hosts and the different infall times of satellites in understanding galaxy quenching.
\end{abstract}


\section{Introduction}
Galaxies at low redshift are sorted out into two populations by their star formation (SF) activity.
One is disk-like `star-forming' galaxies that have ongoing SF so young stellar populations with blue colors are observed.
The other is elliptical-like `quenched (passive)' galaxies whose SF activity is strongly suppressed so old stellar populations with red colors are observed.
Galaxy quenching is closely related to various galaxy properties such as morphology, color, age, star formation rate (SFR), and kinematics \citep[e.g.][and references therein]{Kauffmann+2003,Fang+2013,Brownson+2022}.
Thus, understanding when, where, and how galaxies halt SF is one of the important topics in extragalactic astronomy.

In the observation, galaxies whose stellar mass is larger than $10^{10}$~M$_{\odot}$ show low SFR (specific SFR $\le 10^{-11}$~yr$^{-1}$) regardless of the environment, called `mass quenching' \citep[e.g.,][and references therein]{Peng+2010,Peng+2012,Man+2018}.
In this case, all the internal processes of galaxies (secular evolution) that include outflows from stellar winds, supernova explosion, and active galactic nucleus (AGN) feedback decrease SFR \citep[e.g.,][]{DiMatteo+2005,Wake+2012}.
On the other hand, in a high-density environment the SFR of low-mass galaxies whose stellar mass is smaller than $10^{10}$~M$_{\odot}$ is also suppressed, defined as `environmental quenching' \citep[e.g.,][]{Jaffe+2016,Crossett+2017,Medling+2018,Schaefer+2019,GonzalezDelgado+2022}.
In this case, there are various mechanisms that stop the star formation of galaxies: ram-pressure stripping \citep{Gunn+1972}, tidal stripping\citep{Merritt+1984}, strangulation or starvation \citep{Larson+1980,Balogh+2000}, and harassment \citep{Gallagher+1972}.
These two quenching mechanisms have been consistently suggested to describe the observed properties of passive galaxies \citep[e.g.,][]{Contini+2020,Li+2020}.

In the framework of the Lambda cold dark matter ($\Lambda$CDM) model, it is widely accepted that galaxies are growing hierarchically, which means low-mass galaxies are fundamental building blocks for cosmological structure formation.
In this hierarchical paradigm, low-mass galaxies that fall into high-mass galaxies and/or the larger structures become their satellites, which are expected to be merged eventually.
During their first orbital passage to the pericenter, SF activities of satellites are highly suppressed by ram-pressure stripping and tidal stripping: called `post-processing' \citep[e.g.,][]{Gabor+2010,Vijayaraghavan+2013,Donnari+2021a}.
Interestingly, the SFR suppression of satellites is known to start even before satellites fall into the larger structures, known as `pre-processing' \citep[e.g.,][]{Hou+2014,vanderBurg+2018,Sarron+2019,Sarron+2021}.
Various simulations and observations have been used to study the quenching mechanism of galaxies \citep[e.g.,][and references therein]{Bahe+2013,Bahe+2019,Contini+2020,Rhee+2020,Li+2020,Donnari+2021a,Reeves+2022}, but the relative amount of pre- and post-processings for galaxy quenching is still under discussion.

Since the galaxy groups and clusters are built up with satellites accreted from the outside, the fraction of quenched satellites can be a result that reflects both pre- and post-processings.
In this paper, we use the IllustrisTNG simulation to investigate how the fraction of the quenched satellite in galaxy groups and clusters is determined by the pre- and post-processings.
We also examine the relationship between the fraction of quenched satellites at $z=0$ and various properties of host groups and clusters.

This paper is organised as follows.
In Section \ref{sec2}, we introduce the IllustrisTNG simulations and how we select sample galaxies in our study.
In Section \ref{sec3}, we investigate whether pre- or post-processing is more dominant for galaxy quenching and what makes a diversity of the fraction of quenched galaxies at $z=0$.
Section \ref{sec4} shows which properties of hosts are related to the quenched fractions, and the effect of the passage of the pericenter of hosts on quenching is examined in Section \ref{sec5}.
Section \ref{fig6} summerizes our results.

\section{Method}
\label{sec2} 
\subsection{The IllustrisTNG Simulations}

To examine the quenched fraction of satellites in each galaxy group and cluster, we use the cosmological magnetohydrodynamical simulations of the IllustrisTNG project (hereafter IllustrisTNG)\footnote{\url{http://www.tng-project.org}}, which is composed of three different simulation volumes whose one side length is 50, 100, and 300~Mpc: TNG50, TNG100, and TNG300, respectively \citep{Marinacci+2018,Naiman+2018,Nelson+2018,Nelson+2019a,Pillepich+2018b,Pillepich+2019,Springel+2018}.
To secure the best statistics for the galaxy groups and clusters, we use the TNG300 simulation, which has the largest volume of (300~Mpc)$^3$.
The initial mass resolution of the dark matter (DM) particle and the gas cell is 5.0$\times$10$^{7}$~M$_{\odot}$ and 1.1$\times$10$^{7}$~M$_{\odot}$, respectively.
Detailed baryonic physics bringing the galaxy formation can be found in \citet{Weinberger+2017} and \citet{Pillepich+2018a}.

Halos and subhalos in each snapshot are identified using friends-of-friends \citep[FoF;][]{Davis+1985} and {\sc subfind} \citep{Springel+2001} algorithms.
To find the main progenitor of each halo, we use the merger tree made by {\sc sublink} \citep{Rodriguez-Gomez+2015}.
In this paper, we refer to galaxy groups and clusters as `hosts', and to the galaxies that have been accreted by these hosts as `satellites'.

\subsection{Sample selection}
\label{sec2.2}

\begin{figure*}[ht!]
    \plotone{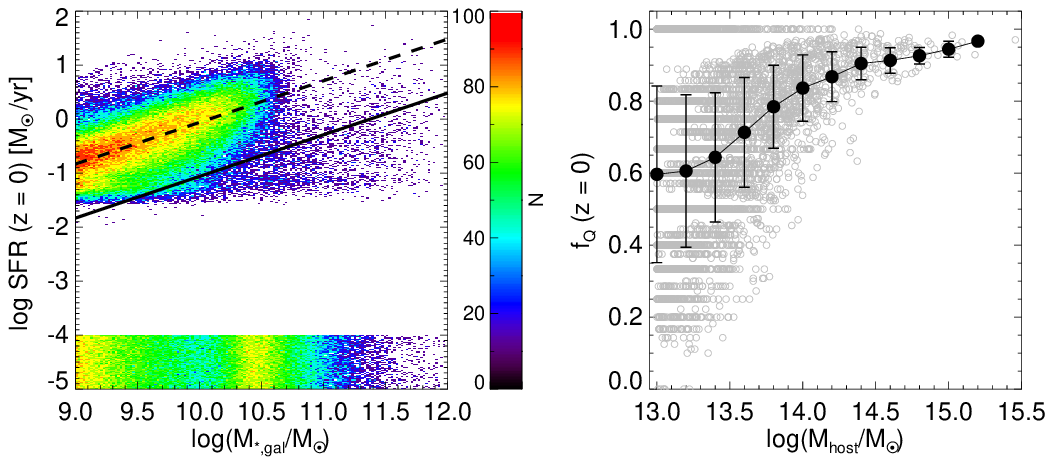}
    \caption
    {
    Left: the SFR as a function of $M_{\rm *,gal}$.
    The black dashed and solid lines represent the star-forming MS and the line 1-dex below from the star-forming MS, respectively.
    The color scale represents the number of galaxies in each pixel.
    We define the `quenched galaxy' whose SFR is below 1 dex from the star-forming MS.
    Galaxies with a zero SFR value are randomly assigned a logarithmic SFR value between -5 and -4 solely to illustrate their existence.
    Right: the number fraction of quenched galaxies to all galaxies in individual groups and clusters ($f_{\rm Q}$) as a function of $M_{\rm host}$ at $z=0$.
    Black solid dots are the mean $f_{\rm Q}$ in each mass bin with an errorbar indicating the standard deviation.
    In the range of low-mass hosts, there are points along the same value of $f_{\rm Q}(z=0)$.
    These low-mass hosts have a small number of satellites ($\sim$4.2 satellites on average) so there is a high probability that the value of $f_{\rm Q}(z=0)$ is discrete.
    }
    \label{fig1}
\end{figure*}

For the completeness of the galaxy samples, we select galaxies whose stellar mass ($M_{\rm *,gal}$) is more massive than $10^{9}$~M$_{\odot}$, which corresponds to more than 100 star particles.
The TNG300 includes 4824 groups (10$^{13}$~M$_{\odot} \le M_{\rm host}<10^{14}$~M$_{\odot}$) and 426 clusters ($M_{\rm host}\ge10^{14}$~M$_{\odot}$), where $M_{\rm host}$ is the virial mass of each host at $z=0$.
The total number of satellites selected as the galaxy samples for the groups and clusters is 40,139 and 25,535, respectively.

We use the SFR and star-forming main sequence (MS) to divide galaxies into star-forming and quenched galaxies, i.e., $\log{\rm~SFR}$ of quenched galaxies is 1 dex below from the star-forming MS \citep{Donnari+2019}.
In this study, the SFR of galaxies is calculated by stars that have formed in recent 200~Myrs \citep{Donnari+2019,Pillepich+2019}.
The star-forming MS is calculated by the following procedure \citep{Donnari+2019,Pillepich+2019}: 1) we calculate the median SFR of star-forming galaxies in 0.2-dex logarithmic bins in the range of $M_{\rm *,gal}=10^{9}$-$10^{10.2}$~M$_{\odot}$, performing a fit to the median SFR linearly, 2) we re-calculate the new median SFR, excluding quenched galaxies, whose SFR is 1 dex below from the previous median SFR, and 3) we repeat the second step until the median SFR converges to a given accuracy (1 per cent).
Finally, we linearly extrapolate the SFR to calculate the median SFR of star-forming galaxies massive than $M_{\rm *,gal}>10^{10.2}$~M$_{\odot}$.
The slope and $y$-intercept used in the linear fitting are 0.77 and -7.77, respectively.

The left panel of Figure \ref{fig1} shows the SFR of all galaxies in TNG300 with $M_{\rm *,gal}$ at $z=0$.
The criterion between star-forming and quenched galaxies is shown as the black solid line, which is 1-dex below from the star-forming MS (the black dashed line).
The right panel of Figure \ref{fig1} shows the fraction of the quenched galaxies to all galaxies in individual hosts ($f_{\rm Q}$, hereafter `the quenched fraction') at $z=0$ as a function of $M_{\rm host}$.
In this panel, the mean $f_{\rm Q}(z=0)$ increases with $M_{\rm host}$, and the scatter in $f_{\rm Q}(z=0)$ decreases as $M_{\rm host}$ increases \citep[e.g.,][]{Donnari+2021a,Donnari+2021b}.
To check whether this trend is an intrinsic scatter or a statistical scatter, we resample the hosts with a bootstrapping resampling in each mass bin while each mass bin has an equal number of hosts.
We still can see the increasing scatter with decreasing $M_{\rm host}$ so this trend is an intrinsic scatter not a statistical scatter from the poor number of high-mass hosts.
The results are consistent even if the quenched galaxies are replaced with red galaxies (see Appendix \ref{appB}).

Note that various papers have shown that the effective kinetic AGN feedback mode quenches most galaxies with $M_{\rm *,gal}>10^{10.3}$~M$_{\odot}$ \citep[e.g.,][]{Nelson+2018,Terrazas+2020,Donnari+2021a}.
In our sample, galaxies massive than $10^{10.3}$~M$_{\odot}$ are mostly quenched by strong AGN feedback.
To check how AGN feedback affects the results in Figure \ref{fig1}, we examine the relation between $f_{\rm Q}(z=0)$ and $M_{\rm host}$, using low-mass satellites ($M_{\rm *,gal}<10^{10.3}$~M$_{\odot}$).
Although there is an offset of the median $f_{\rm Q}(z=0)$ of low-mass satellites ($\sim$0.1~dex lower than that of all satellites), we find that the trend of the median $f_{\rm Q}(z=0)$ of low-mass satellites is similar to that of all satellites: 1) $f_{\rm Q}(z=0)$ is increasing with $M_{\rm host}$, and 2) the size of errorbars is decreasing with $M_{\rm host}$.


\section{Host-Mass Dependency and the Scatter of Quenched Fractions}
\label{sec3}

We investigate why the mean $f_{\rm Q}(z=0)$ increases with $M_{\rm host}$ by dividing hosts into three mass ranges: low-mass ($10^{13.0}$~M$_{\odot}<M_{\rm host}\le10^{13.5}$~M$_{\odot}$), intermediate-mass ($10^{13.5}$~M$_{\odot}<M_{\rm host}\le10^{14.0}$~M$_{\odot}$), and high-mass ($10^{14.0}$~M$_{\odot}\le M_{\rm host}$) bins or hosts.
We then examine the origin of a scatter in $f_{\rm Q}(z=0)$.

\subsection{The Effects of Pre- and Post-Processings on the Quenched Fraction}
\label{sec3.1}

\begin{figure*}[ht!]
    \includegraphics[width=\textwidth]{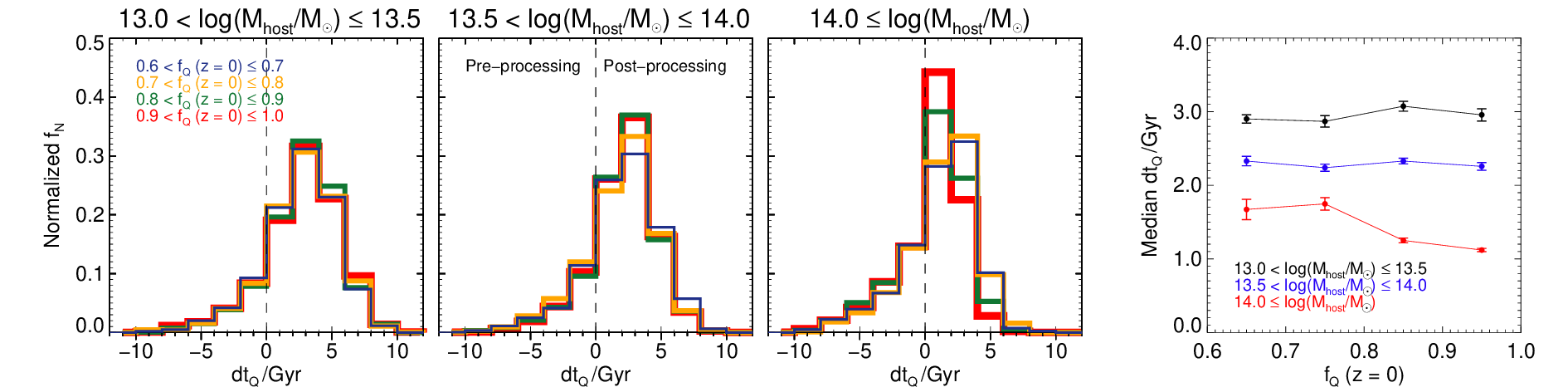}
    \caption
    {
    Left: the normalized number of satellites with the time interval (${\rm d}t_{\rm Q}$) from when it first falls into the host until it is quenched.
    In each $M_{\rm host}$ bin, we divide satellites according to $f_{\rm Q}(z=0)$ ranging from 0.6 to 1.0 with a 0.1 interval (blue, orange, green, and red lines).
    The grey dashed lines are the infall time of satellites.
    Right: the median ${\rm d}t_{\rm Q}$ of satellites with $f_{\rm Q}(z=0)$ in each $M_{\rm host}$ bin.
    Errorbars indicate the standard error.
    }
    \label{fig2}
\end{figure*}

\begin{figure*}[ht!]
    \includegraphics[width=\textwidth]{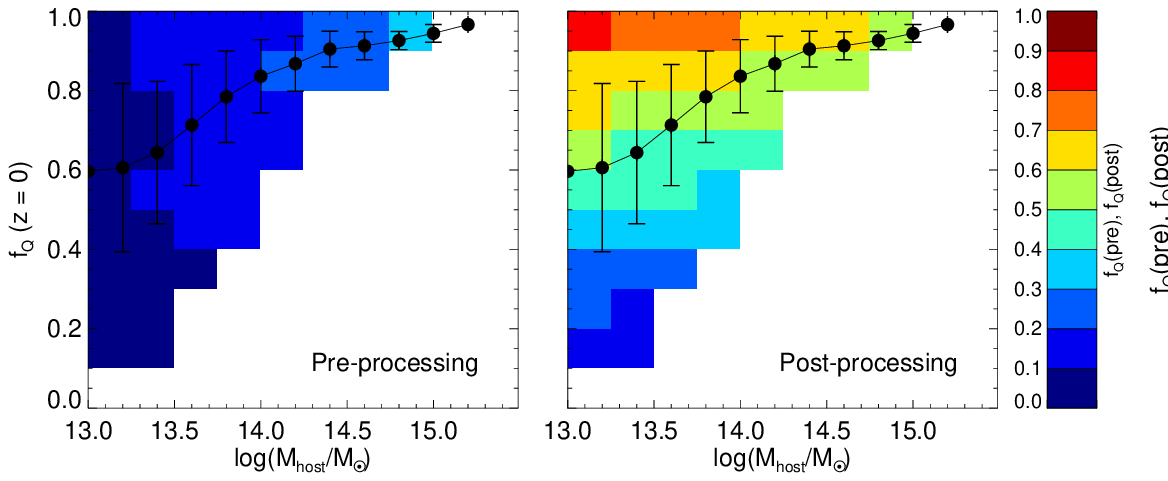}
    \caption
    {
    Left: the 2D histograms showing the distribution of the mean $f_{\rm Q}(z=0)$ and $M_{\rm host}$.
    The color indicates the mean $f_{\rm Q}$(pre) of satellites in each pixel before satellites fall into their hosts.
    Black solid dots with 1$\sigma$ error bars are the same as in the right panel of Figure \ref{fig1}.
    Middle: same as the left panel but the color denotes $f_{\rm Q}$(post).
    Right: $f_{\rm Q}$(pre) and $f_{\rm Q}$(post) as a function of $f_{\rm Q}(z=0)$ in each mass bin.
    }
    \label{fig3}
\end{figure*}

In the right panel of Figure \ref{fig1}, the mean $f_{\rm Q}(z=0)$ is increasing with $M_{\rm host}$: a host-mass dependency of $f_{\rm Q}(z=0)$, shown in various observations \citep[e.g.,][]{Muzzin+2013,Jian+2018,Fang+2018,Donnari+2021b,Reeves+2022}.
To investigate the origin of the host mass-dependency of $f_{\rm Q}(z=0)$, we measure the time interval (${\rm d}t_{\rm Q}$) of each satellite from when it first falls into the host until it is quenched.
Quenched satellites are defined by the MS at each snapshot and we define the time of quenching/infall as the time of the first snapshot when galaxies are identified as quenched/satellites.

The left three panels of Figure \ref{fig2} show the normalized number of galaxies ($f_{\rm N}$) with ${\rm d}t_{\rm Q}$ for hosts in three different mass bins.
Here, the negative ${\rm d}t_{\rm Q}$ indicates that the satellites have been quenched before infall, while the positive ${\rm d}t_{\rm Q}$ indicates quenching after infall.
Regardless of $M_{\rm host}$, the number of satellites that are quenched after falling into their final hosts is dominant (${\rm d}t_{\rm Q}>0$): the number fraction of quenched galaxies after infall is 0.84, 0.81, and 0.70 for low-, intermediate-, and high-mass bins, respectively.
It means pre-processing may have contributed to reducing the SF but not fully quenched satellites but post-processing mainly contributes to the galaxy quenching \citep[e.g.,][]{Donnari+2021a}.
The decreasing number fraction of quenched fractions after infall for more massive hosts indicates that there are more pre-processed satellites in the more massive hosts \citep[e.g.,][]{Hou+2014,vanderBurg+2018,Bahe+2019}.
Because satellites in more massive hosts can have more massive infalling groups than less massive hosts \citep[e.g.,][]{Hou+2014,vanderBurg+2018,Sarron+2019,Donnari+2021b}, satellites are more likely to be quenched than those in the low-mass hosts \citep[e.g.,][]{Reeves+2022,Salerno+2022}.
Indeed, among total quenched satellites, 25.8\% of those in low-mass hosts and 50.2\% of those in high-mass hosts are members of other structures when they fall into the final low- and high-mass hosts \citep[e.g.,][]{Han+2018,Donnari+2021b}.

Note that these results can be affected by the high-mass satellites that experience strong AGN feedback.
When we measure the normalized $f_{\rm N}$ of low-mass satellites ($M_{\rm *,sat}<10^{10.3}$~M$_{\odot}$), most of them are quenched by post-processing not pre-processing. 
It means most pre-processed satellites are quenched by strong AGN feedback in our sample \citep[e.g.,][]{Donnari+2021a}.
However, the results with low-mass satellites are still similar to those with all satellites 1) the number of post-processed satellites is more than that of pre-processed satellites, and 2) the median ${\rm d}t_{\rm Q}$ is decreasing with increasing $M_{\rm host}$.

The other thing that we have to notice in the left three panels of Figure \ref{fig2} is the peak ${\rm d}t_{\rm Q}$ is getting shorter as $M_{\rm host}$ is increasing.
To measure this trend quantitatively, we plot the median ${\rm d}t_{\rm Q}$ with $f_{\rm Q}(z=0)$ (the right panel of Figure \ref{fig2}).
In high-mass hosts, which generally have denser and more extended gas reservoirs (intracluster/intergalactic medium), ram-pressure stripping is more efficient and affects infalling satellites earlier, making satellites quenched faster compared to those in low-mass hosts \citep[e.g.,][]{Jaffe+2015,Boselli+2016}.
As a result, satellites in high-mass hosts undergo more intense pre- and post-processings than those in low-mass hosts.
It enables high-mass hosts to have higher $f_{\rm Q}(z=0)$ than low-mass hosts, as shown in the right panel of Figure \ref{fig1}.

Interestingly, the median ${\rm d}t_{\rm Q}$ is steeply decreasing with $f_{\rm Q}(z=0)$ in a high $M_{\rm host}$ bin (the red line in the right panel of Figure \ref{fig2}).
This trend is coming from the fact that the high-mass bin covers the wide range of $M_{\rm host}$ from $10^{14}$~M$_{\odot}$ up to $\sim10^{15.5}$~M$_{\odot}$.
Indeed, among high-mass hosts, hosts with $0.8<f_{\rm Q}(z=0)\le1.0$ are more massive than other hosts with $0.6<f_{\rm Q}(z=0)\le0.8$ on average.
Because more massive hosts might experience the most efficient pre- and post-processings, satellites in more massive hosts are more rapidly quenched than those in less massive hosts.

In Figure \ref{fig2}, we demonstrate why high-mass hosts have higher $f_{\rm Q}(z=0)$ than low-mass hosts: satellites in high-mass hosts experience pre-processing most strongly (0.30, low- and intermediate-hosts are 0.16 and 0.19, respectively) and are rapidly quenched by efficient post-processing.
However, Figure \ref{fig2} does not clearly show whether pre- or post-processings contribute to the scatter of $f_{\rm Q}(z=0)$ as shown in the right panel of Figure \ref{fig1}.
To investigate this, we separately examine the fraction of pre- and post-processings and plot the right two panels in Figure \ref{fig3}, which shows the effects of $f_{\rm Q}$(pre) and $f_{\rm Q}$(post) on $f_{\rm Q}(z=0)$.
The fractions of $f_{\rm Q}$(pre) and $f_{\rm Q}$(post) are calculated by taking an average of the amount of pre- and post-processed satellites in each host and bins that have more than 10 data are used for the statistical significance.
In the left panel of Figure \ref{fig3}, the mean $f_{\rm Q}$(pre) only depends on $M_{\rm host}$ but not depends on $f_{\rm Q}(z=0)$ so only the host-mass dependency of $f_{\rm Q}$(pre) is shown (see also Figure \ref{fig2}).
It demonstrates that satellites are slightly quenched by pre-processing that depends on $M_{\rm host}$, before they fall into their final hosts.
However, in the middle panel of Figure \ref{fig3}, the mean $f_{\rm Q}$(post) depends not only on $M_{\rm host}$ but also on$f_{\rm Q}(z=0)$, which demonstrates post-processing makes most satellites quenched and the large scatter in $f_{\rm Q}(z=0)$ of low-mass hosts.

To measure the amount of pre- and post-processed satellites quantitatively, we plot $f_{\rm Q}$(pre) and $f_{\rm Q}$(post) as a function of $f_{\rm Q}(z=0)$ in the right panel in Figure \ref{fig3}.
We take an average of $f_{\rm Q}$(pre) and $f_{\rm Q}$(post) in each $f_{\rm Q}(z=0)$ bin so that $f_{\rm Q}(z=0)=f_{\rm Q}$(pre)$+f_{\rm Q}$(post).
Satellites in high-mass hosts have slightly higher $f_{\rm Q}$(pre) than those in low- and intermediate-mass hosts, indicating that pre-processing is more effective for them \citep[e.g.,][]{Donnari+2021a}.
However, $f_{\rm Q}$(pre) is almost constant with $f_{\rm Q}(z=0)$ at a $M_{\rm host}$, which means pre-precessing is strongly depends on $M_{\rm host}$ not $f_{\rm Q}(z=0)$.
On the other hand, $f_{\rm Q}$(post) increases with decreasing $M_{\rm host}$ because satellites in low-mass hosts are less quenched by pre-processing than those in intermediate- and high-mass hosts.
In addition, the mean $f_{\rm Q}(z=0)$ at a fixed $M_{\rm host}$ strongly depends on $f_{\rm Q}$(post).
Therefore, although satellites are slightly quenched by pre-processing that depends on $M_{\rm host}$, post-processing that makes most satellites quenched mainly cause the scatter of $f_{\rm Q}(z=0)$.

\subsection{The Effect of the Infall Time of Satellites on the Quenched Fraction}
\label{sec3.2}

\begin{figure*}[ht!]
    \includegraphics[width=\textwidth]{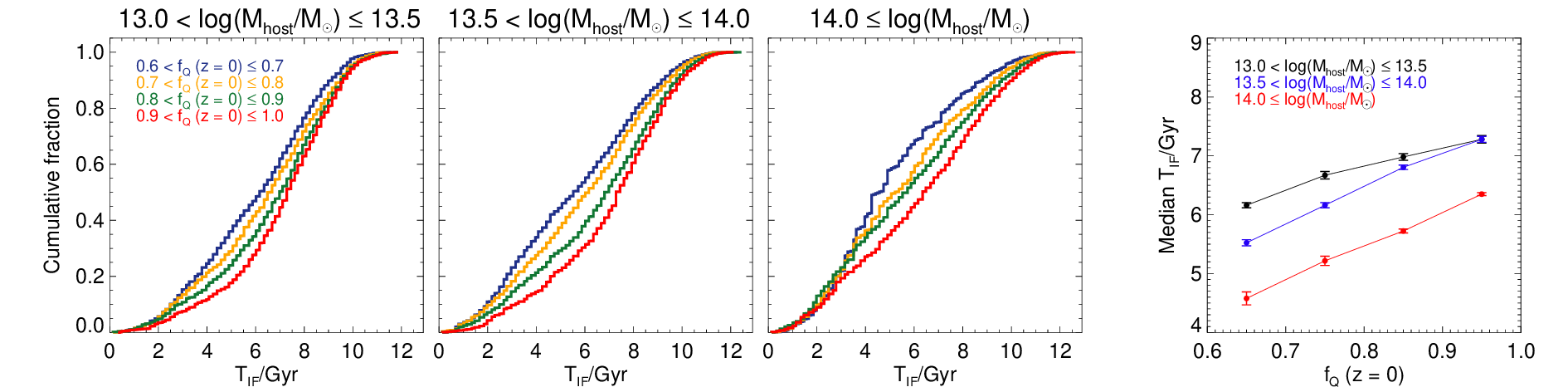}
    \caption
    {
    Left: the cumulative fraction of satellites as a function of the infall time of satellites ($T_{\rm IF}$)  when satellites fall into their hosts.
    Right: the median $T_{\rm IF}$ as a function of $f_{\rm Q}(z=0)$ in each mass bin with an errorbar indicating the standard error.
    }
    \label{fig4}
\end{figure*}

\begin{figure*}[ht!]
    \includegraphics[width=\textwidth]{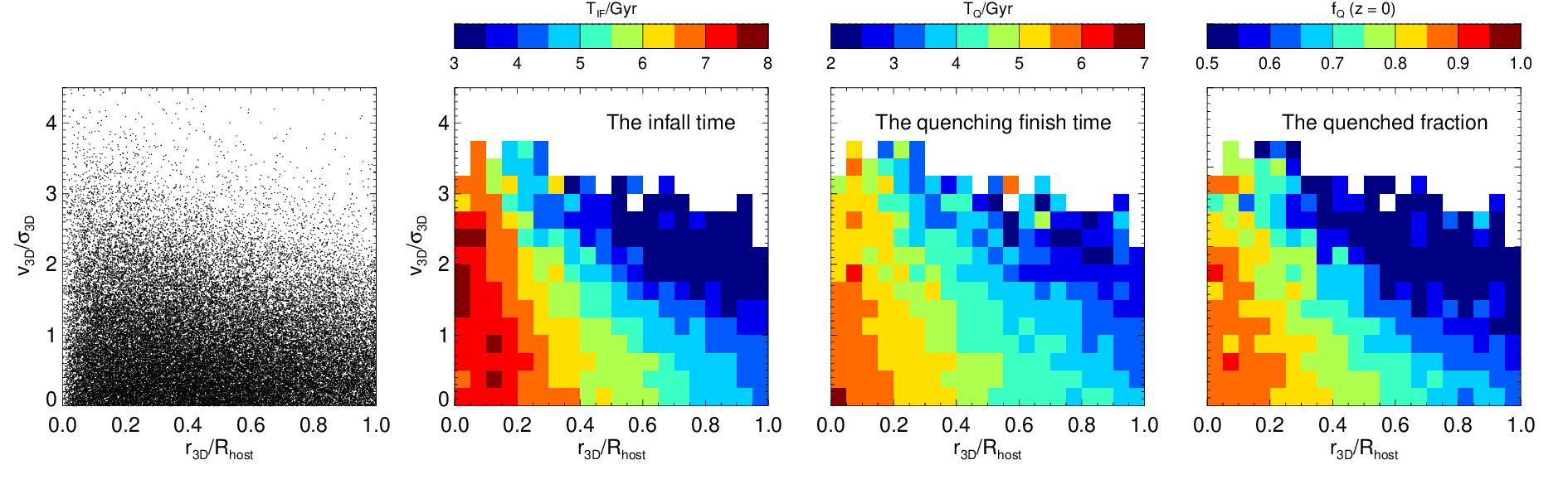}
    \caption
    {
    Distributions in the 3D phase-space diagram of satellites (the left penal), where $R_{\rm host}$ is a viral radius of hosts.
    The 3D phase-space diagram for satellites, where colors indicate the mean $T_{\rm IF}$ (the second panel), the mean quenching finish time ($T_{\rm Q}$; the third panel), and the mean $f_{\rm Q}(z=0)$ (the forth panel) in each pixel of the grid.
    }
    \label{fig5}
\end{figure*}

\begin{figure*}[ht!]
    \includegraphics[width=\textwidth]{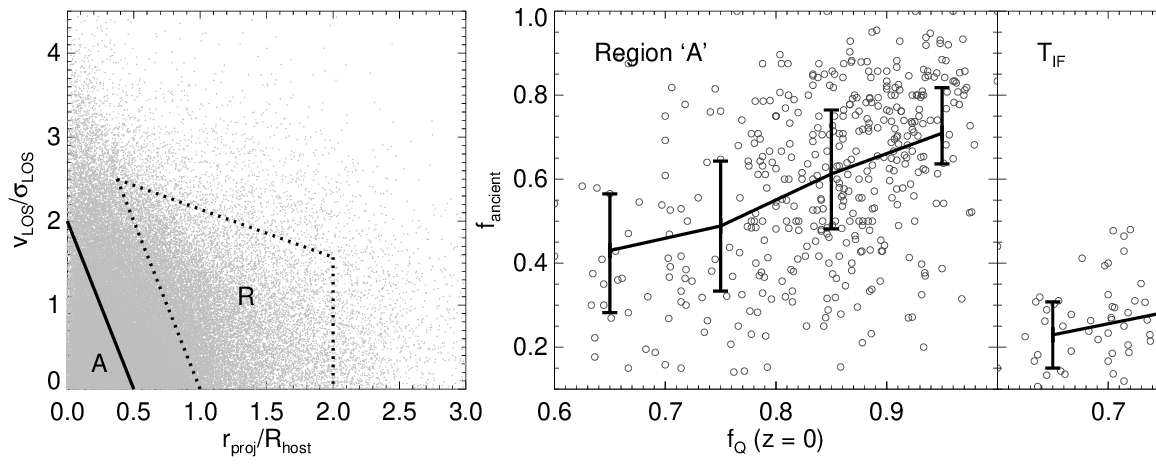}
    \caption
    {
    Left: the regions on the projected phase-space diagram where we use to calculate $f_{\rm ancient}$ (see the text in detail), where gray dots are the distribution of satellites.
    Middle: the relation between $f_{\rm Q}(z=0)$ and $f_{\rm ancient}$ of hosts massive than $M_{\rm host}\ge10^{14}$~M$_{\odot}$.
    Ancient infallers are defined as satellites in the region `A'.
    The black solid line connects the median values and the errorbars show the first and the third quartiles.
    Right: symbols are the same as in the middle panel, but ancient infallers are defined as satellites that fall into their hosts 6.64~Gyrs ago.
    }
    \label{fig6}
\end{figure*}

In Figure \ref{fig3}, we showed that post-processing makes the scatter of $f_{\rm Q}(z=0)$: there is a diversity of $f_{\rm Q}(z=0)$ values in a narrow range of $M_{\rm host}$.
In this section, we focus on the factors that drive the diversity $f_{\rm Q}(z=0)$ values.
As shown in Figures \ref{fig2} and \ref{fig3}, post-processing plays an important role in galaxy quenching so the time when satellites fall into their hosts and how long satellites stay in their hosts can have a significant impact on galaxy quenching \citep[e.g.,][]{Smith+2019}.
To investigate this further, we define the infall time ($T_{\rm IF}$) as the time when satellites first fall into their final hosts, and measure it as a lookback time.
A large median $T_{\rm IF}$ in a given $M_{\rm host}$ bin indicates that hosts gain their satellites early, while the small median $T_{\rm IF}$ represents hosts gain their satellites recently.
Thus, the median $T_{\rm IF}$ indicates the mass growth history of hosts within a similar range of $M_{\rm host}$.

The left three panels of Figure \ref{fig4} show the cumulative fraction of satellites with $T_{\rm IF}$, indicating that satellites in hosts with high $f_{\rm Q}(z=0)$ fall into their hosts early regardless of $M_{\rm host}$.
These early-infall satellites stay in their hosts long enough so most of these satellites are quenched by post-processing.
To measure this trend quantitatively, we plot the median $T_{\rm IF}$ with $f_{\rm Q}(z=0)$ in the right panel of Figure \ref{fig4}.
The median $T_{\rm IF}$ increases with $f_{\rm Q}(z=0)$, supporting the result that a large number of satellites that fall into their hosts early are more likely to be quenched by post-processing.
In other words, satellites have a high probability of being quenched as they stay longer in their hosts.
Furthermore, at a fixed median $T_{\rm IF}$, high-mass hosts have the highest $f_{\rm Q}(z=0)$ because satellites in high-mass hosts experience pre-processing most strongly and are rapidly quenched by post-processing (as shown in Section \ref{sec3.1}).

The early-infall satellites lose their orbital energy due to the gravitational drag force and gradually move towards the center of hosts, known as the `dynamical friction' \citep{Chandrasekhar1943}.
As a result, early-infall satellites tend to stay close to the center of their hosts, losing their orbital energy, while late-infall satellites are more likely to be located in the outskirts of their hosts.
This spatial separation between early- and late-infall satellites is shown in a phase-space diagram, which is a useful tool for studying the relationship between the infall time of satellites and their location in hosts \citep[e.g.,][]{Oman+2016,Rhee+2017,Reeves+2022}.
To investigate how $T_{\rm IF}$ and $f_{\rm Q}(z=0)$ of satellites are shown in a phase-space diagram, we plot the left panel of Figure \ref{fig5}.
Satellites that fall into their final hosts early are located innermost region, indicating that these satellites might have spent most of their time in their final hosts.
We also plot the quenching finish time ($T_{\rm Q}$) of satellites in a phase-space diagram to compare it with $T_{\rm IF}$.
We define $T_{\rm Q}$ as $T_{\rm IF}-{\rm d}t_{\rm Q}$, which represents the lookback time when the quenching of satellites is completed.
The similarity in color gradient between $T_{\rm IF}$ and $T_{\rm Q}$ indicates that satellites that fall into their hosts earlier tend to be quenched earlier by post-processing \citep[e.g.,][]{Smith+2019}.
Finally, we plot the right panel of Figure \ref{fig5} to examine the relationship of $T_{\rm IF}$ and $T_{\rm Q}$ to $f_{\rm Q}(z=0)$.
It shows that satellites located innermost region are the most quenched.
The good agreement in color gradient of the three panels in Figure \ref{fig5} shows that satellites that fall into their final hosts early could be most quenched because they have enough time in their hosts.

In Figure \ref{fig4}, we show that hosts with a higher $f_{\rm Q}(z=0)$ have a higher median $T_{\rm IF}$ than those with a lower $f_{\rm Q}(z=0)$.
To compare this result with the observation, we can use an observable projected phase-space diagram that can estimate the mean $T_{\rm IF}$ (Jeong et al. in prep.).
Based on the results in Figure \ref{fig5}, satellites can be divided into early- and late-infall satellites in a projected phase-space diagram \citep[see also][]{Oman+2016,Rhee+2017}.
To define early- and late-infall satellites, we measure the projected radius ($r_{\rm proj}$) and the line-of-sight velocity ($v_{\rm LOS}$) of each satellite.
For the projected radius, we take an average of the radius from the center of hosts to their satellites in each $xy-$, $yz-$, and $xz-$ plane.
For the line-of-sight velocity, we correct the Hubble flow \citep[][see their equation (2)]{Oman+2016} and take an average of $v_{\rm LOS}$ in each $xy-$, $yz-$, and $xz-$ plane.
The left panel of Figure \ref{fig6} shows early- and late-infall satellites defined as `ancient infallers (A)' and `recent infallers (R)' \citep[e.g.,][]{Rhee+2017,Jeong+2019}\footnote{Note that we follow the method in \citet{Jeong+2019} who divide the phase-space diagram into two regions for statistical significance unlikely \citet{Rhee+2017} who divide it into five regions.}.
We define the region `A', which includes most ancient infallers that fall into the final hosts before $6.45$~Gyr and define the region `R', which includes most first infallers not fallen yet into their final hosts but will fall into their final hosts \citep[see their table~2 and figure~6][]{Rhee+2017}.

The middle panel of Figure \ref{fig6} shows the fraction of ancient infallers, $f_{\rm ancient}=A/(A+R)$, with $f_{\rm Q}(z=0)$.
As $f_{\rm Q}(z=0)$ increases, $f_{\rm ancient}$ also increases, which indicates that hosts with high $f_{\rm Q}(z=0)$ have many ancient infallers so hosts gain their mass early.
Thus, ancient infallers are likely to be more quenched by post-processing because they have spent most of their time in their final hosts.

Since we define ancient infallers as satellites within the region `A', which is determined by projected position and velocity, the projection effect may cause some non-member satellites to be included in the region `A'.
To check the projection effect of the observation in the middle panel, we alternatively define `ancient infallers' as satellites that fall into their hosts 6.45~Gyrs ago \citep[][see their Table~2]{Rhee+2017}, using the $T_{\rm IF}$ instead of the region `A'.
The right panel of Figure \ref{fig6} shows the results.
When we use $T_{\rm IF}$, which can be measured in simulations, the value of $f_{\rm ancient}$ is much smaller than the value in the middle panel because, among the ancient infallers in the middle panel, about 30\% of them fall into their hosts after 6.45 Gyr.
This implies that a large number of recent infallers are misidentified as ancient infallers due to projection effects.
However, both panels show that hosts with high $f_{\rm Q}(z=0)$ tend to have a large population of ancient infallers, which suggests that they have grown earlier compared to hosts with low $f_{\rm Q}(z=0)$.
The trend of $f_{\rm ancient}$ that is measured from a phase-space diagram (the middle panel) is similar to that of $f_{\rm ancient}$ that is measured from $T_{\rm IF}$ in simulations (the right panel).
Thus, we can estimate the mass growth of hosts from the observable value of $f_{\rm ancient}$.

Note that high-mass galaxies ($10^{10.5}$~M$_{\odot}<M_{\rm *,gal}<10^{11.0}$~M$_{\odot}$) in IllustrisTNG are overquenched by strong baryonic feedback from AGN \citep{Nelson+2018}.
Overquenching makes quenched fractions high compared to observations \citep[e.g.,][]{Sherman+2020,Angthopo+2021} so there might be some offset of $f_{\rm Q}(z=0)$ between simulations and observations.
However, we still can compare hosts with more early growth to those with more recent growth by using the relative values of $f_{\rm Q}(z=0)$.

\section{Quenched fractions with Various Properties of Hosts}
\label{sec4}

\begin{figure*}[ht!]
    \plotone{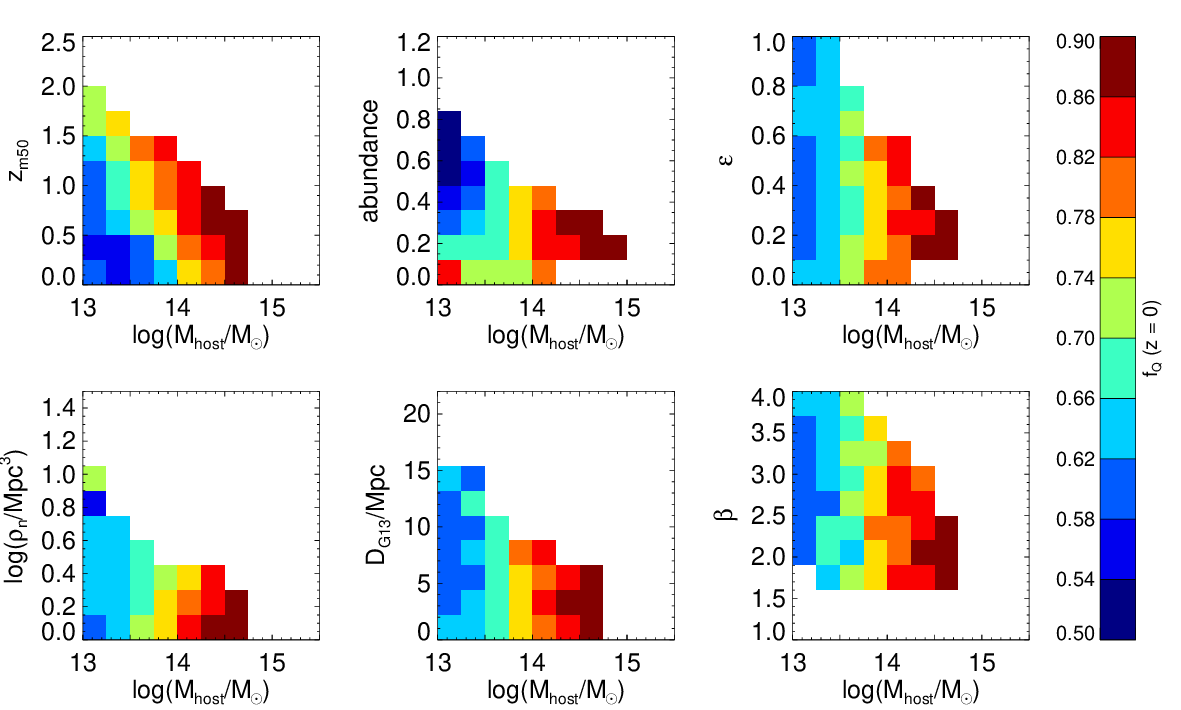}
    \caption
    {
    The 2D histograms of various properties of hosts with $M_{\rm host}$: the host formation time ($z_{\rm m50}$), abundance, ellipticity ($\epsilon$), number density ($\rho_{\rm n}$), nearest distance to groups or clusters ($D_{\rm G13}$), and isotropy ($\beta$).
    Color scales are $f_{\rm Q}(z=0)$ at $z=0$.
    For statistical reliability, we use pixels that have more than 10 data points.
    }
    \label{fig7}
\end{figure*}

\begin{figure}[ht!]
    \plotone{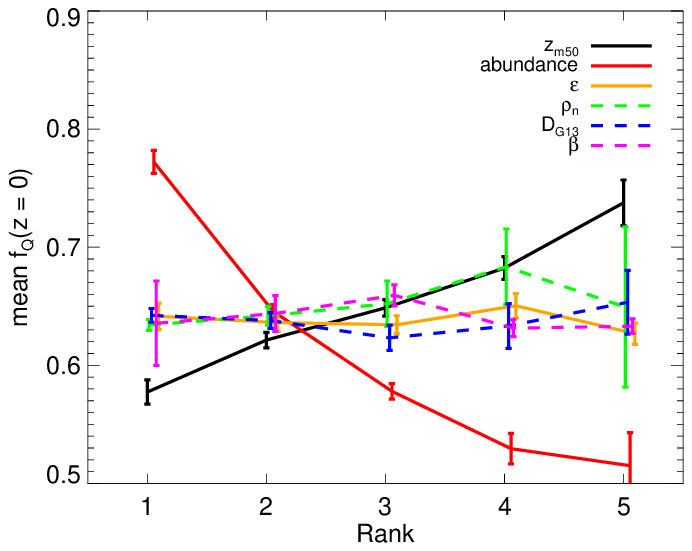}
    \caption
    {
    The relation between $f_{\rm Q}(z=0)$ and the rank of various properties of hosts (see the text in detail).
    Data in Figure \ref{fig7} is used to calculate ranks.
    }
    \label{fig8}
\end{figure}

In this section, we investigate various properties of hosts that are related to their mass growth history and $f_{\rm Q}(z=0)$.
There are various observational properties of galaxy clusters: the abundance of satellites, density, colors, morphology, ellipticity, isotropy, and so on.
Among them, we use the abundance of satellites, ellipticity ($\epsilon$), number density ($\rho_{\rm n}$), nearest distance to groups or clusters ($D_{\rm G13}$), isotropy of satellites ($\beta$).
Satellites whose stellat mass ($M_{\rm *,sat}$) is larger than $10^{9}$~M$_{\odot}$ are used to calculate each property.
The cluster formation time ($z_{\rm m50}$), which is the redshift when hosts first reach 50\% of $M_{\rm host}(z=0)$, is included as an indicator of the mass growth history of hosts \citep[e.g.,][]{Chun+2023}.
Unlikely other observational properties, $z_{\rm m50}$ is not an observable property.
However, we have already investigated the different $T_{\rm IF}$ can affect the scatter of $f_{\rm Q}(z=0)$ so how $z_{\rm m50}$ can affect $f_{\rm Q}(z=0)$ has to be examined.

We define the abundance of satellites as below:
\begin{equation}
    \frac{N_{\rm sat}}{M_{\rm host}}\times10^{12},
\end{equation}
where $N_{\rm sat}$ is the number of satellites in the virial radius ($R_{\rm host}$).
We use the abundance normalized by $M_{\rm host}$ because in general, $N_{\rm sat}$ depends strongly on $M_{\rm host}$ \citep[e.g.,][see their left panel of figure~5]{Wu+2022}.
Thus, if the abundance is small in a given range of $M_{\rm host}$, hosts are dynamically more evolved than hosts with a large abundance.

We calculate the total ellipticity ($\epsilon$) using the positions of satellites relative to the hosts \citep[][see their equations~(5)-(8) for details]{Shin+2018}.
We measure the ellipticity of satellites in $R_{\rm host}$ in the $xy$-, $yz$-, and $xz$-planes, and take an average of these ellipticities ($\epsilon$).
Thus, $\epsilon$ denotes how satellites inside the host distribute.

To measure the number density of satellites around each host ($\rho_{\rm n}$), we count the number of satellites from $R_{\rm host}$ to $3R_{\rm host}$.
We calculate the nearest distance ($D_{\rm G13}$) to any group or cluster whose $M_{\rm host}$ is larger than $10^{13}$~M$_{\odot}$.
Thus, hosts with small $D_{\rm G13}$ are located in denser environments, while those with larger $D_{\rm G13}$ are in less dense regions.

We measure the isotropy of satellites from 0 to 5$R_{\rm host}$, using the azimuthal symmetric excess \citep [$\beta$:][]{Gouin+2022} with harmonic orders of $m=1$ to $m=4$:
\begin{equation}
    \beta=\sum_{m=1}^{n}\beta_{m}=\sum_{m=1}^{n}\frac{|Q_{m}|}{|Q_{0}|},
\end{equation}
where $|Q_{m}|$ is the modulus of the aperture multipole moment at the order of $m$.
In $xy$-,$yz$-, and $xz$-planes, we compute $\beta$ and take an average of these three $\beta$.

Figure \ref{fig7} shows 2D histograms of $z_{\rm m50}$ and various observational properties of hosts with $M_{\rm host}$.
The properties of hosts and $M_{\rm host}$ are divided into 10 bins to examine their variations.
We find that $f_{\rm Q}(z=0)$ strongly depends on $M_{\rm host}$ (vertical stripes of colors) and weakly depends on $z_{\rm m50}$ (diagonal stripes of colors) and the abundance of satellites (only when $M_{\rm host}$ is small).
In Sections \ref{sec3.1} and \ref{sec3.2}, we find out that post-processing is more important for galaxy quenching than
pre-processing; that is, satellites falling into their hosts early are likely to be more quenched due to a longer stay in their hosts compared to those falling later.
Hosts with a high $z_{\rm m50}$ have many early-infall satellites (a high mean $T_{\rm IF}$), while those with a low $z_{\rm m50}$ have many late-infall satellites (a low mean $T_{\rm IF}$).
Therefore, the different mass growth history results in diverse values of $f_{\rm Q}(z=0)$. 

On the other hand, in low-mass hosts, $f_{\rm Q}(z=0)$ decreases with increasing the abundance of satellites.
In a given range of $M_{\rm host}$, hosts are dynamically unrelaxed if the abundance of satellites is high, indicating that satellites have fallen recently.
These satellites have not stayed in their hosts long enough to be quenched so eventually hosts could have low $f_{\rm Q}(z=0)$.
Other properties of hosts, $\epsilon$, $\rho_{n}$, $D_{\rm G13}$, and $\beta$, do not show the relation to $f_{\rm Q}(z=0)$ because those properties are related to the environment or the mass distribution rather than the mass growth history of hosts.

To quantitatively measure the variation of $f_{\rm Q}(z=0)$ with various properties of hosts in Figure \ref{fig7}, we plot Figure \ref{fig8}.
Only low-mass hosts ($10^{13}$~M$_{\odot}<M_{\rm host}<10^{13.5}$~M$_{\odot}$) are used for statistical reliability.
We first sort the values of each property in ascending order and then divide each property into 5 ranks of 0-20\%, 20-40\%, 40-60\%, 60-80\%, and 80-100\%, i.e., rank 1 represents the lowest value of each property and rank 5 denotes the highest value.
Next, we calculate the mean $f_{\rm Q}(z=0)$ of each rank.
In Figure \ref{fig8}, the mean $f_{\rm Q}(z=0)$ decreases with the abundance of satellites and increases with $z_{\rm m50}$.
Hosts with a high $z_{\rm m50}$ gain their mass early due to the early infall of their satellites.
Early-infall satellites have spent more time in the potential of their hosts on average, and naturally, lose more mass as they are subjected to post-processing, which can affect not only the mass loss of satellites but also the destruction of satellites.
After satellites fall into their hosts, they experience hydrodynamical interaction with the intracluster medium (ICM) \citep[e.g.,][]{Gunn+1972}, ram-pressure stripping, so satellites can halt the SF activity.
Satellites are also affected by gravitational tides in the potential well of their hosts so dark matter, stars and gas in satellites can be stripped \citep[e.g.,][]{Merritt+1984}.
Harassment \citep[e.g.,][]{Moore+1996} and galaxy mergers \citep[e.g.,][]{Toomre+1972} also can distort and destroy satellites.
Some satellites lose their orbital energy and finally merge into the cluster galaxy (Brightest Cluster Galaxy), dynamical friction \citep{Chandrasekhar1943}.
As a result, the destruction rate of early-infall satellites is higher, making the abundance of satellites lower.
Therefore, in a low $M_{\rm host}$ bin, the abundance of satellites and the fraction of quenched satellites at $z=0$ can be observable indicators of the mass growth history of hosts.

We find that in intermediate- and high-mass hosts, only $z_{\rm m50}$ has a relation to $f_{\rm Q}(z=0)$.
The dependence of $T_{\rm IF}$ with $f_{\rm Q}$ has already been shown (the right panel of Figure \ref{fig4}) so we can still find this trend when it comes to $z_{\rm m50}$.
Satellites in intermediate- and high-mass hosts fall into their hosts more recently and are quenched more rapidly than those in low-mass hosts (the right panels of Figures \ref{fig2} and \ref{fig4}).
However, these satellites are not disrupted easily although they fall into their hosts a long time ago \citep[e.g.,][]{Bahe+2019}.
It makes the dependence of the abundance on $f_{\rm Q}$ weak unlikely in the case of low-mass hosts.
Therefore, only the fraction of quenched satellites at $z=0$ can be used as an indicator of the mass growth history of intermediate- and high-mass hosts.

\section{Effect of Host Pericenter Passage on Quenched Satellites}
\label{sec5}

\begin{figure*}[ht!]
    \plotone{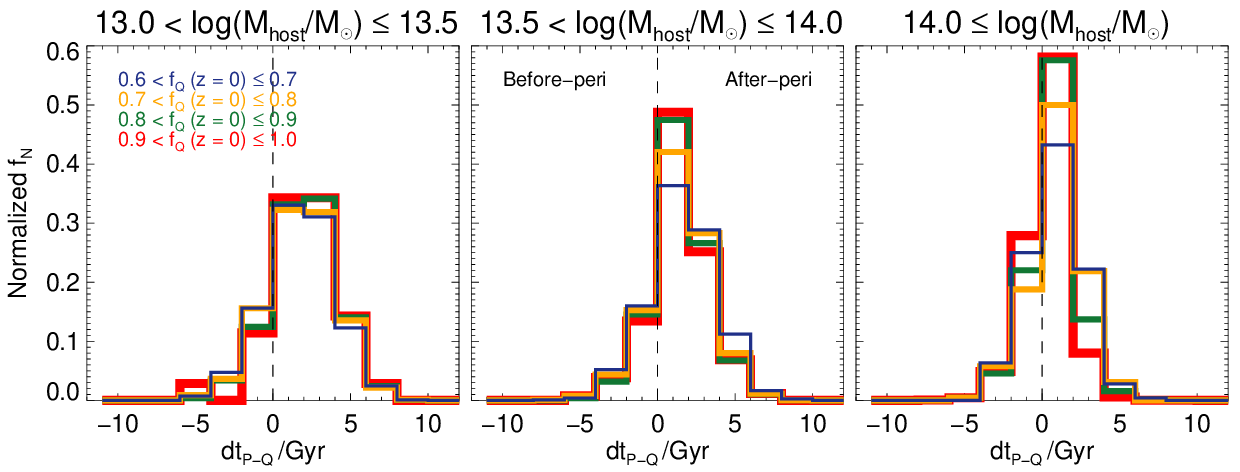}
    \caption
    {
    The normalized number of satellites with the time interval (${\rm d}t_{\rm P-Q}$) from the time when satellites pass by the pericenter to the quenching finish time in different $M_{\rm host}$ bins.
    In each mass bin, we divide satellites according to $f_{\rm Q}(z=0)$ ranging from 0.6 to 1.0 with a 0.1 interval (blue, orange, green, and red lines).
    The grey dashed lines denote the time when satellites are passing by the pericenter.
    }
    \label{fig9}
\end{figure*}


In this paper, we emphasize the importance of post-processing for satellites to be quenched because 1) most satellites are quenched after they fall into their final hosts, and 2) satellites that fall into their hosts are quenched by efficient post-processing so the different mean $T_{\rm IF}$ of satellites makes a scatter of $f_{\rm Q}(z=0)$.
Although satellites can be quenched by some physical mechanisms, we are unable to differentiate between them.
Instead, we can derive the effect of the passage of the pericenter for galaxy quenching.
Various studies have shown that galaxies are quenched around the first pericenter \citep[e.g.,][]{Rhee+2020,Upadhyay+2021}.
\citet{Upadhyay+2021} analyze a sample of 11 massive ellipticals in the Coma cluster.
They find that the SF in those ellipticals stops 1~Gyr after the first pericenter passage and this rapid quenching comes from the combination of ram-pressure stripping with tidal interactions.
To investigate the effect of the passage of the pericenter of hosts on galaxy quenching, we trace the orbit of each satellite and measure the time interval from when satellites pass by the pericenter ($T_{\rm peri}$) to the quenching finish time: ${\rm d}t_{\rm P-Q} = T_{\rm peri}-T_{\rm Q}$.

Figure \ref{fig9} shows the normalized number of satellites with ${\rm d}t_{\rm P-Q}$.
Note that we exclude satellites that are pre-processed or do not pass by the pericenter to see the effect of the pericenter.
Satellites with ${\rm d}t_{\rm P-Q}>0$ are quenched after passing by the pericenter, while those with ${\rm d}t_{\rm P-Q}<0$ are quenched before.
Most satellites are quenched after passing by the pericenter so the interaction between the passage of the pericenter and satellites is important for galaxy quenching.
We find that high-mass hosts have a slightly higher quenched fraction before the pericenter (0.31) compared to low- and intermediate-mass hosts (0.21 and 0.22, respectively).
Although we remove satellites that do not pass by the pericenter or are pre-processed, some satellites in high-mass hosts might have low SFRs possibly approaching zero when they fall due to efficient pre-processing (see Section \ref{sec3.1}) and post-processing upon falling into their hosts, before the passage of the pericenter.
This is why high-mass hosts have large quenched fractions before the passage of the pericenter.
The mean ${\rm d}t_{\rm P-Q}$ is 1.9, 1.4, and 0.6~Gyrs in low-, intermediate-, and high-mass bin, respectively.
The shortest mean ${\rm d}t_{\rm P-Q}$ in high-mass hosts might result from the efficient ram-pressure stripping \citep[e.g.,][]{Boselli+2016,Oman+2016} and tidal stripping \citep[e.g.,][]{Upadhyay+2021}.

We also measure ${\rm d}t_{\rm P-Q}$ with $M_{\rm *,sat}$ to investigate how satellites are affected by the passage of the pericenter.
Because high-mass satellites ($10^{10.3}$~M$_{\odot}< M_{\rm *,sat}$) are affected by strong AGN feedback, we only investigate ${\rm d}t_{\rm P-Q}$ of low-mass satellites ($10^{9.5}\le~M_{\rm *,sat}<10^{10.3}$).
In the case of low-mass satellites, the average ${\rm d}t_{\rm Q}$ and ${\rm d}t_{\rm P-Q}$ is $\sim$3.1~Gyr and $\sim$1.4~Gyr, respectively, so low-mass satellites pass by the pericenter of their hosts after $\sim$1.7~Gyr since infall.
It suggests that the SFR of low-mass satellites is rarely affected after the first infall or the SFR is gradually suppressed via ram-pressure stripping or strangulation \citep[e.g.,][]{Wetzel+2013,Taranu+2014,Haines+2015,Foltz+2018,Maier+2019,Roberts+2019,Rhee+2020}.
When low-mass satellites fall into their final hosts, the density of the ICM around the outskirt of hosts is low so cold gas in satellites cannot interact with the ICM around them.
During this phase, steady gas depletion (starvation/strangulation) occurs in low-mass satellites.
Thus, they can continue to increase their $M_{\rm *,sat}$ because their SFR is not completely suppressed.
When satellites approach the inner region of their hosts, they can reach the threshold ICM density ($\sim10^{-28}$~g~cm$^{-3}$), and then, a significant fraction of cold gas becomes susceptible to ram-pressure stripping.
Finally, satellites can be quenched after the passage of the pericenter of their hosts \citep[e.g.,][]{Wetzel+2013,Roberts+2019,Rhee+2020}.
The other thing that we have to pay attention to is that ${\rm d}t_{\rm P-Q}$ of low-mass satellites is increasing with $M_{\rm *,sat}$.
If $M_{\rm *,sat}$ decreases, the potential well is getting shallower so gases in low-mass satellites are likely to be stripped by tides or ram pressure \citep{Jung+2018,Rohr+2023}.

In this section, we investigate how the passage of the pericenter can affect the SF of satellites.
We find that satellites in high-mass hosts are slightly more quenched than those in low- and intermediate-hosts due to efficient pre- and post-processings.
After passing by the pericenter of high-mass hosts, satellites are most rapidly quenched due to the efficient ram-pressure stripping.
Low-mass satellites are rapidly quenched, passing by the pericenter after a delay phase.
The time from the passage of the pericenter to being quenched increases with the stellar mass of satellites because gases in low-mass satellites might be easily stripped due to a shallow potential well.

\section{Summary}
\label{sec6}
We investigate the fraction of quenched satellites of hosts whose virial mass is larger than $10^{13}$~M$_{\odot}$ at $z=0$, using TNG300 in the IllustrisTNG cosmological magnetohydrodynamical simulations.
In simulations, the fraction of quenched satellites depends on the virial mass of hosts and there is a scatter of quenched fractions at $z=0$.
Throughout the paper, we examine which physical mechanisms cause those results and which properties of hosts are related to the fraction of quenched satellites.
Belows are a summary of our results.
\begin{enumerate}
    \item Post-processing is important for satellites to be quenched because 1) most satellites are quenched after falling into their hosts and 2) the fraction of quenched satellites at $z=0$ is increasing with the fraction of quenched satellites after infall.
    
    \item Satellites in high-mass hosts experience pre-processing more than those in low- and intermediate-mass hosts and they are rapidly quenched by efficient post-processing after falling into their hosts.
    It makes satellites in high-mass hosts most quenched at $z=0$ so there is a host-mass dependency of the fraction of quenched satellites at $z=0$.
    
    \item The different infall time of satellites makes a scatter in the fraction of quenched satellites at $z=0$ in a given mass range of hosts.
    Satellites that fall into their hosts early have spent most of their time in their hosts so they could be more quenched than those falling lately.
    Thus, hosts with a high fraction of quenched satellites at $z=0$ have many early-infall satellites, while hosts with low quenched fractions have many late-infall satellites.
    
    \item In a phase-space diagram, satellites that fall into their hosts early are located in the innermost region of hosts.
    These satellites have spent most of their time in their hosts, and are most quenched by post-processing.
    Thus, if hosts have a high fraction of quenched satellites at $z=0$, the fraction of early-infall satellites is high, which means hosts gain their mass early.
    It indicates that if we know the fraction of quenched satellites at $z=0$, we can estimate the mass growth history of hosts.
    
    \item Among the various properties of hosts, $z_{\rm m50}$ and the abundance of satellites are related to the fraction of quenched satellites at $z=0$ in low-mass hosts ($10^{13}$~M$_{\odot}<M_{\rm host}\le10^{13.5}$~M$_{\odot}$).
    Thus, the abundance and the fraction of quenched satellites at $z=0$ can be indicators of the history of hosts indirectly in low-mass hosts.
    In intermediate- and high-mass hosts, only $z_{\rm m50}$ is related to the fraction of quenched satellites at $z=0$.

    \item Most satellites are quenched after passing by the pericenter of their hosts.
    The time interval from when satellites pass by the pericenter to the quenching finish time decreases with the mass of hosts because of effective ram-pressure stripping and tidal stripping in massive hosts.
    Furthermore, low-mass satellites seem to experience a delay phase before quenching because most of them spend more than half of their time since infall to the passage of the pericenter and then are quenched after the passage of the pericenter.

\end{enumerate}

\begin{acknowledgments}
This work was supported by the Korea Astronomy and Space Science Institute under the R\&D program (Project No. 2023-1-830-01) supervised by the Ministry of Science and ICT (MSIT).
KWC was supported by the National Research Foundation of Korea (NRF) grant funded by the Korea government (MSIT) (2021R1F1A1045622).
H.J. acknowledges support from the National Research Foundation of Korea (NRF) grant funded by the Korea government (MSIT) (No. NRF-2019R1F1A1041086).
JHL acknowledges support from the National Research Foundation of Korea (NRF) grant funded by the Korea government (MSIT) (No. 2022R1A2C1004025).
This research was partially supported by the Australian Research Council Centre of Excellence for All Sky Astrophysics in 3 Dimensions (ASTRO 3D), through project number CE170100013.
\end{acknowledgments}

\appendix

\section{Plots using the red galaxy fraction}
\label{appB}
\restartappendixnumbering

\begin{figure}[ht!]
    \plotone{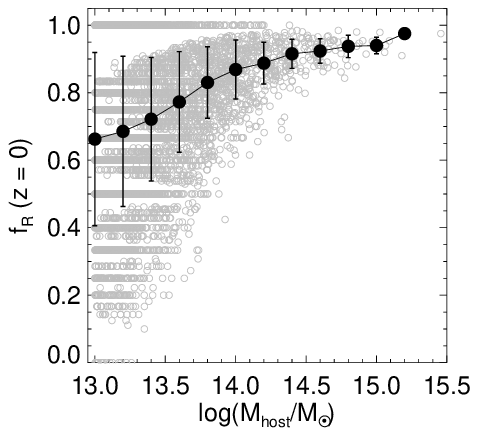}
    \caption
    {
    The number fraction of red galaxies ($f_{\rm R}$) to all galaxies in individual groups and clusters as a function of $M_{\rm host}$ at $z=0$.
    Black solid dots are the mean $f_{\rm R}$ in each mass bin with 1$\sigma$ errors.
    }
    \label{figb1}
\end{figure}

\begin{figure*}[ht!]
    \plotone{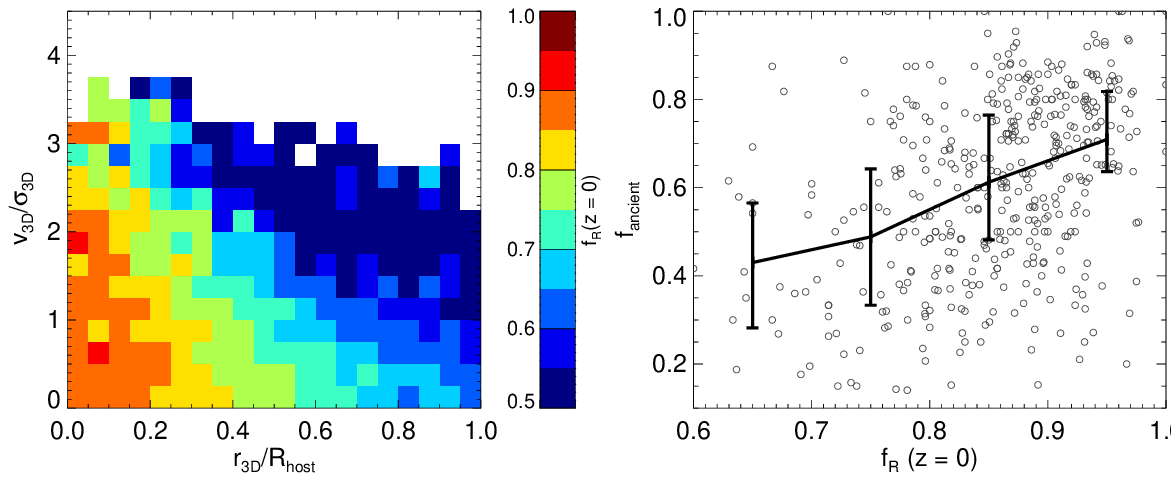}
    \caption
    {
    Left: the mean $f_{\rm R}(z=0)$ in a phase-space diagram.
    Right: $f_{\rm ancient}$ with $f_{\rm R}(z=0)$.
    }
    \label{figb2}
\end{figure*}

\begin{figure}[ht!]
    \plotone{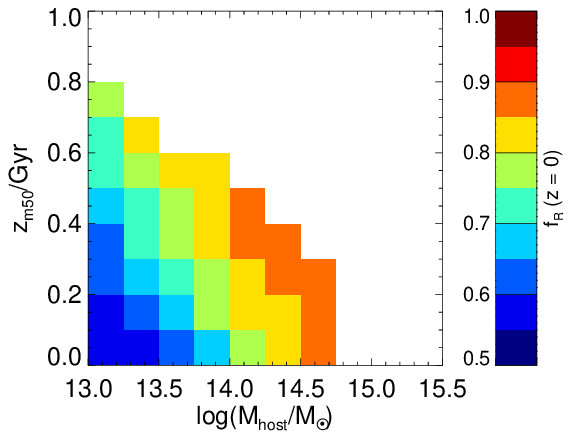}
    \caption
    {
    The 2D histogram of $z_{\rm m50}$ with $M_{\rm host}$.
    The color bar represents $f_{\rm R}(z=0)$.
    Pixels that have more than 10 data are used for statistical reliability.
    }
    \label{figb3}
\end{figure}

The Red galaxy in TNG300 is defined as galaxies that have the $g-r$ color, larger than $(g-r) = 0.5\log{M_{\rm *,gal}}+0.1$~mag \citep[][see their equation~(6)]{Pulsoni+2020}.
We plot Figure \ref{figb1}, which shows the red galaxy fraction ($f_{\rm R}$) with $M_{\rm host}$ at $z=0$.
The trend of $f_{\rm R}(z=0)$ increases with $M_{\rm host}$, similar to the trend in the right panel of Figure \ref{fig1}.

We plot the mean $f_{\rm R}(z=0)$ in a phase-space diagram (the left panel of Figure \ref{figb2}).
The trend of color gradient is similar to that in the right panel of Figure \ref{fig5}.
After ancient infallers are defined as satellites located in the region `A' (see the left panel of Figure \ref{fig6}), we plot the fraction of ancient infallers with $f_{\rm R}(z=0)$ in the right panel of Figure \ref{figb2}.
The fraction of ancient infallers increases with $f_{\rm R}(z=0)$, indicating that satellites falling into their hosts early become red galaxies more than those falling recently.

In Figure \ref{figb3}, we investigate the relation between $z_{\rm m50}$ and $f_{\rm R}(z=0)$.
Similar to the upper left panel of Figure \ref{fig7}, $f_{\rm R}(z=0)$ also can be an indicator that represents the mass growth history of hosts.

\bibliography{red_galaxy}{}
\bibliographystyle{aasjournal}



\end{document}